\documentclass[preprint,prd,aps,tighten,nofootinbib,amssymb]{revtex4}
\usepackage{graphicx}
\usepackage{epsfig,amssymb,amsmath}

\newcommand{\beq}{\begin{equation}}
\newcommand{\eeq}{\end{equation}}
\newcommand{\bea}{\begin{eqnarray}}
\newcommand{\eea}{\end{eqnarray}}
\newcommand{\bear}{\begin{array}}
\newcommand {\eear}{\end{array}}
\newcommand{\bef}{\begin{figure}}
\newcommand {\eef}{\end{figure}}
\newcommand{\bec}{\begin{center}}
\newcommand {\eec}{\end{center}}

\def\GEV#1{10^{#1}{\rm\,GeV}}

\begin{document}
\draft
\tighten
\preprint{TU-938, IPMU13-0114}
\title{\large \bf
Self-interacting Dark Radiation
}
\author{
    Kwang Sik Jeong\,$^a$\footnote{email: ksjeong@tuhep.phys.tohoku.ac.jp}
    and
    Fuminobu Takahashi\,$^{a,b}$\footnote{email: fumi@tuhep.phys.tohoku.ac.jp}}
\affiliation{
 $^a$ Department of Physics, Tohoku University, Sendai 980-8578, Japan\\
 $^b$ Kavli IPMU, TODIAS, University of Tokyo, Kashiwa 277-8583, Japan
    }

\vspace{2cm}

\begin{abstract}
We consider a simple class of models where dark radiation has self-interactions and therefore
does not free stream.  Such dark radiation has no anisotropic stress (or  viscosity), leaving
a distinct signature on the CMB angular power spectrum.
Specifically we study a possibility that hidden gauge bosons and/or chiral fermions
account for the excess of the effective number of neutrino species.
They have gauge interactions and remain light due to the unbroken hidden gauge symmetry,
leading to $\Delta N_{\rm eff} \simeq 0.29$ in some case.
\end{abstract}

\maketitle

The Planck satellite measured the temperature anisotropies of the cosmic microwave background
(CMB) with unprecedented precision. The baryon acoustic oscillations are induced by the balance
between the gravity and the photon pressure, and the shape of the CMB angular power
spectrum is sensitive to the energy content of the Universe and its properties.
In particular, the Planck results
constrained the amount of relativistic degrees of freedom as~\cite{Ade:2013zuv}
\beq
N_{\rm eff} = 3.30^{+0.54}_{-0.51},
\label{NeffPl}
\eeq
at $95\%$C.L. where the Planck, WMAP polarization, ground-based CMB measurements
and Baryon Acoustic Oscillation data are used. Compared to the standard model (SM) value,
$N_{\rm eff}^{\rm SM} = 3.046$, there is a mild preference for the excess
of the effective number of neutrinos, $\Delta N_{\rm eff} = N_{\rm eff} - N_{\rm eff}^{\rm SM}
\simeq 0.30 >0$, coined dark radiation.

Let us introduce new light degrees of freedom to account for the excess.
Then there are two questions that immediately arise~\cite{Nakayama:2010vs}:
\begin{itemize}
\item Why do they remain relativistic until the recombination epoch?
\item Why is their abundance  about one third of a single neutrino species,
and why not larger or smaller?
\end{itemize}
The answer to the first question depends on the production process.
If the new particles are produced by the decay of heavy fields,\footnote{For the early works,
see e.g. Ref.~\cite{Ichikawa:2007jv}, in which non-thermal production of dark radiation as well as
its connection to (warm) dark matter were considered. There are many recent ones along the similar lines,
which are omitted here due to the limited space.}
they may remain relativistic at late times. However, it is non-trivial to explain the observed abundance,
and it sometimes leads to the overproduction of dark radiation~\cite{Higaki:2013lra}.

On the other hand, if the new particles were once in thermal equilibrium with the SM particles,
they must be extremely light
in order to remain relativistic until the recombination epoch.
The required ultra-light mass strongly suggests the existence of symmetry forbidding mass terms.
Such symmetry can be  (i) gauge symmetry,  (ii) chiral symmetry,  or (iii) shift symmetry, for which
gauge bosons, chiral fermions and Nambu-Goldstone (NG) bosons are
a candidate for dark radiation, respectively.
The effective number of neutrino species receives a contribution,
\begin{equation}
\label{dnf}
    \Delta N_{\rm eff}
    = \left(\frac{8}{7} N_{g} + N_f + \frac{4}{7} N_{\rm GB} \right)
 \left( \frac{g_{*\nu}}{g_{*{\rm dec}}} \right)^{4/3},
\end{equation}
from massless $N_g$ gauge bosons, $N_f$ chiral fermions and $N_{\rm GB}$ NG bosons
in a hidden sector, which are assumed to decouple at the same temperature.
Here $g_{*\nu} (=10.75)$ and $g_{*{\rm dec}}$ denote the relativistic degrees of freedom
in the SM sector evaluated at the time of decoupling of neutrinos and the hidden
sector fields, respectively.
Interestingly, the excess $\Delta N_{\rm eff}$ can be naturally of order unity in this case.
These possibilities were studied in Ref.~\cite{Nakayama:2010vs}.\footnote{
Recently the possibility of NG bosons as dark radiation was revisited in Ref.~\cite{Weinberg:2013kea},
where the SM fields are assumed to be singlet under the symmetry, and the connection of the
symmetry to the dark matter was considered.
See also Ref.~\cite{Baek:2013qwa}.
}

We assume throughout this letter  that the dark radiation was once in thermal equilibrium with the SM sector
and it decoupled some time before the big bang nucleosynthesis (BBN),
along the lines of Ref.~\cite{Nakayama:2010vs}.  Then the next question is,
\begin{itemize}
\item How do we distinguish between these candidates for dark radiation?
\end{itemize}
Since the dark radiation is assumed to be decoupled from the SM plasma before BBN,
it is generically difficult to distinguish between different models.
As we shall see, however, the dark radiation can have self-interactions in the cases (i) and (ii),
if the  symmetry remains unbroken. Then the dark radiation does not free stream and so there is no anisotropic
stress or the viscosity of dark radiation, in contrast to what is usually assumed.
(Note that the bound (\ref{NeffPl}) is for non-interacting dark radiation.)
Such self-interacting dark
radiation should have left a distinct signature in the CMB angular power spectrum.

Let us mention the related works in the past. The self-interacting dark radiation
was considered in a context of neutrinos interacting through majoron exchange~\cite{Chacko:2003dt}.\footnote{
More recently, dark radiation interacting  with dark matter was studied in
Refs. \cite{Archidiacono:2013lva, Franca:2013zxa}.
}
In the present work, the neutrinos have only the usual weak interactions, while it is dark radiation
that is self-interacting, and therefore its effect on the CMB is considered to be milder.
The observational bound on such possibility was studied in e.g.
Refs.~\cite{Bell:2005dr,Cirelli:2006kt,Friedland:2007vv,Smith:2011es,Archidiacono:2011gq,Diamanti:2012tg} before the Planck.
As of writing,  there is no study on the self-interacting dark radiation using the Planck data,\footnote{
In Ref.~\cite{Gerbino:2013ova}, the viscosity parameter of the neutrinos was studied using the Planck data,
while $N_{\rm eff}$ was set to be the SM value.
}
but according to Ref.~\cite{Friedland:2007vv}, the Planck  should be slightly more sensitive to
such self-interacting dark radiation compared to the non-interacting case.
Although it is not certain whether the latest observation favors the self-interacting dark radiation or not,
we expect that $\Delta N_{\rm eff} \sim 0.3 \ll N^{\rm SM}_{\rm eff}$ is  allowed considering
the success of the standard $\Lambda$CDM model and the expected sensitivity.
To be concrete, therefore, we adopt $\Delta N_{\rm eff} \sim 0.3$ as a reference value
in the case of self-interacting dark radiation.

\vspace{5mm}

In this work we introduce a hidden sector with unbroken gauge group $G^\prime$, and
study if it can account for dark radiation with $\Delta N_{\rm eff} \simeq 0.3$.
A simple possibility is to consider a hidden Abelian gauge symmetry U(1)$^\prime$.
Then there is kinetic mixing with the hypercharge field strength $B_{\mu \nu}$,
\beq
\label{kin-mix}
{\cal L} = -\frac{1}{4} F_{\mu \nu}' F'^{\mu \nu}+\kappa F_{\mu \nu}' B^{\mu \nu},
\eeq
where $F'_{\mu\nu} = \partial_\mu A'_\nu - \partial_\nu A'_\mu$, and $\kappa$ parameterizes
the kinetic mixing.
For unbroken U(1)$^\prime$, the above kinetic mixing can be absorbed by field redefinition.
Thus hidden matter fields charged under U(1)$^\prime$ are needed to thermalize the hidden sector.
However if one adds only chiral fermions, there are tight constraints on the kinetic mixing,
$\kappa \lesssim 10^{-15}$~\cite{Davidson:2000hf}, which makes it impossible to thermalize
the hidden sector fields.

A connection between the hidden and visible sector can naturally be achieved by introducing
a SM singlet scalar field $\phi$ charged under $G^\prime$ because the symmetry allows for
a renormalizable coupling,
\bea
\frac{\lambda}{4} |\phi|^2|H|^2,
\label{quartic}
\eea
where $H$ is the SM Higgs doublet field.
The discussion below does not assume a particular hidden gauge group, and we will return
to the Abelian case later.
Let us consider the case that $\phi$ does not develop a vacuum expectation value so that
the hidden gauge group remains unbroken, and that $\phi$ is heavier than the weak scale
to suppress invisible Higgs decay into hidden gauge bosons and scalars.
The hidden gauge bosons interact with the SM sector through the effective coupling,
\bea
\label{eff-c1}
{\cal L}_{\rm eff} = \frac{1}{\Lambda^2_\phi} F^\prime_{\mu\nu} F^{\prime \mu\nu}
|H|^2,
\eea
which is induced by the loop of $\phi$. Here the gauge indices are omitted.
At temperatures below the mass of $\phi$, the scale $\Lambda_\phi$ is estimated to be
\bea
\Lambda_\phi \sim \left(\frac{\lambda g^{\prime 2} }{8\pi^2}\right)^{-1/2}m_\phi,
\eea
dropping a coefficient of order unity,
where $g^\prime$ is the hidden gauge coupling, and $m_\phi$ is the mass of $\phi$.
The hidden gauge bosons interact also with the SM fermions,
\bea
\label{eff-c2}
{\cal L}_{\rm eff} =
\frac{1}{\Lambda^2_\phi} \frac{m_f}{m^2_h}
F^\prime_{\mu\nu} F^{\prime \mu\nu} \bar f f,
\eea
after electroweak symmetry breaking.
Here $f$ denotes the SM fermion with mass $m_f$, and $m_h\simeq 125.5$ GeV
is the mass of the Higgs boson $h$.
Note that there is no mixing between the Higgs boson and the hidden scalar
for $\langle \phi \rangle=0$.

The electroweak symmetry breaking leads to the decay of the Higgs boson
into hidden gauge bosons through the coupling (\ref{eff-c1}).
If the branching ratio is larger than about $1\%$, such invisible Higgs decay can be
probed at future collider experiments such as ILC \cite{Simon:2012ik}.
The decay rate of $h$ into hidden gauge bosons reads
\bea
\Gamma_{h\to V^\prime V^\prime} = \frac{N_g v^2 m_h^3}{2\pi \Lambda_\phi^4}
\simeq  1 {\rm MeV}\times N_g
\left(\frac{1.7\times 10^3{\rm GeV}}{\Lambda_\phi}\right)^4,
\eea
where $v=\langle|H^0|\rangle \simeq 174$ GeV, while the decay rate of $h$ into SM particles
is given by $\Gamma_{h\to {\rm SM}}\simeq 4$ MeV.
The global fit analysis excluded a Higgs branching fraction greater than $19\%$ into
invisible particles at 95\% C.L.~\cite{Giardino:2013bma}, which requires
$\Gamma_{h\to V^\prime V^\prime} \lesssim 1$\,MeV. To satisfy the experimental bound, we need
$\Lambda_\phi$ higher than about 1.7 TeV for $N_g=1$.
The lower bound on $\Lambda_\phi$ can be translated into the constraint on $m_\phi$, $\lambda$,
and $g^\prime$.

On the other hand, if $\phi$ is lighter than half of the Higgs boson mass, $m_\phi<m_h/2$,
$h$ directly decays into hidden scalars through the coupling (\ref{quartic}).
In order for the branching fraction not to exceed $19\%$, the coupling $\lambda$ should
be smaller than about $10^{-2}$.

\begin{figure}[t]
\begin{center}
\begin{minipage}{16.4cm}
\centerline{
{\hspace*{0cm}\epsfig{figure=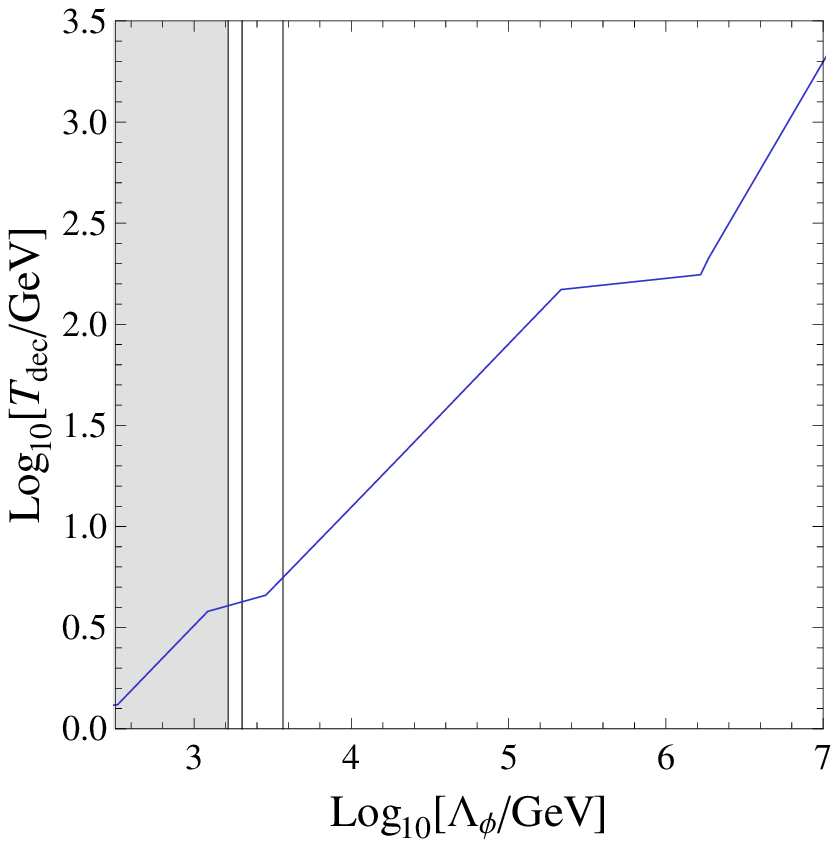,angle=0,width=7.2cm}}
{\hspace*{.4cm}\epsfig{figure=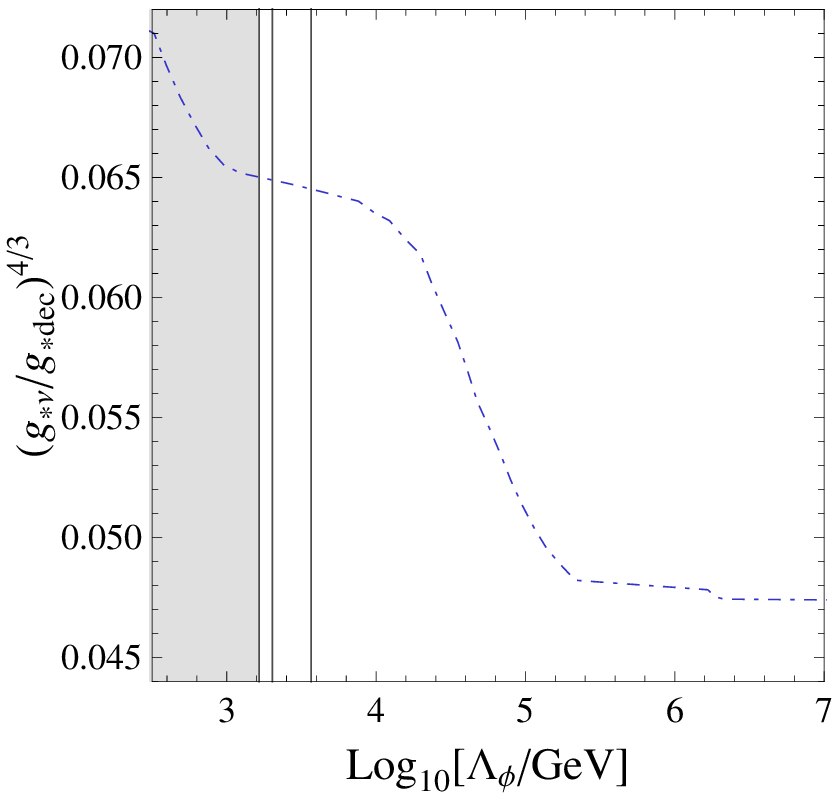,angle=0,width=7.2cm}}
}
\caption{
Dependence of $T_{\rm dec}$ (left panel) and $(g_{\ast \nu}/g_{\ast{\rm dec}})^{4/3}$
(right panel) on $\Lambda_\phi$.
The Higgs decay rate into hidden gauge bosons is proportional to $N_g/\Lambda^4_\phi$,
and it is larger than 1 MeV for $N_g=1$ in the shaded region.
The branching ratio of this mode is ${\rm Br}(h\to V^\prime V^\prime)=20,\,10,\,1\%$
on the vertical lines, from left to right, for the case with $N_g=1$ and $2m_\phi>m_h$.
}
\label{fig:g}
\end{minipage}
\end{center}
\end{figure}

The hidden sector fields are thermalized through the scalar interaction (\ref{quartic}) and the hidden
gauge interactions at sufficiently high temperatures, unless the couplings are extremely small.
At temperatures below the mass of $\phi$,  the effective couplings (\ref{eff-c1}) and (\ref{eff-c2}) can
keep the hidden sector fields in thermal equilibrium with the SM particles.
A rough estimate gives the decoupling temperature of the hidden gauge boson to be
\bea
T_{\rm dec} \sim 200{\rm GeV}
\left(\frac{g_\ast}{100}\right)^{1/6}
\left(\frac{\Lambda_\phi}{10^6{\rm GeV}}\right)^{4/3},
\eea
for $\Lambda_\phi$ larger than about $10^6$ GeV.
Here $g_\ast$ counts all the relativistic degrees of freedom including those in the hidden sector.
For smaller values of $\Lambda_\phi$, the effective couplings to the SM fermions
play an important role.
For instance, the decay temperature reads
\bea
T_{\rm dec} \sim 4.2{\rm GeV}
\left(\frac{g_\ast}{80}\right)^{1/10}
\left(\frac{m_f}{4.2{\rm GeV}}\right)^{-2/5}
\left(\frac{\Lambda_\phi}{2.5\times 10^3 {\rm GeV}}\right)^{4/5},
\eea
if $\Lambda_\phi$ lies in the range between about $2.5\times 10^3$~GeV and
$2.6\times 10^5$~GeV, for which $f$ is considered to be the bottom quark since it is the heaviest
of the SM fermions that are in thermal equilibrium at the decoupling.
Using the property that the effective coupling to $\bar f f$ is proportional to $m_f$, one
finds that the decoupling temperature cannot be lower than the charm quark mass unless
$\Lambda_\phi$ is much smaller than the weak scale.
In the left panel of Fig.~\ref{fig:g}, we show the decoupling temperature as a function of $\Lambda_\phi$.
The plateau around $\Lambda_\phi = \GEV{6}$ corresponds to a case where
the hidden gauge bosons are decoupled when the Higgs boson disappears from the plasma.

\begin{figure}[t]
\begin{center}
\begin{minipage}{16.4cm}
\centerline{
{\hspace*{0cm}\epsfig{figure=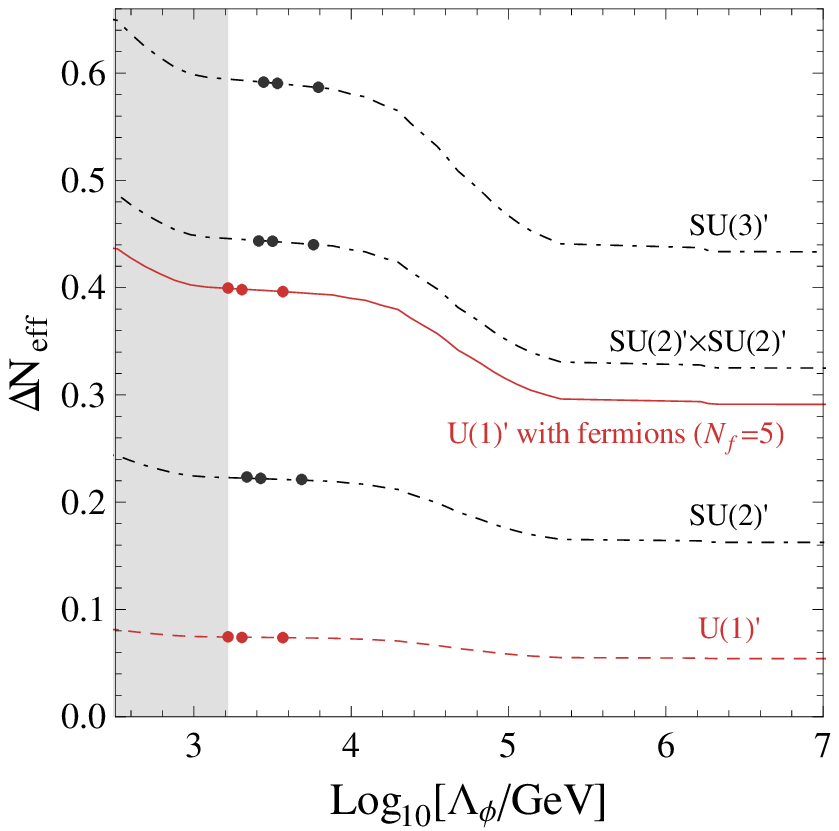,angle=0,width=8.6cm}}
}
\caption{
Dark radiation from the hidden sector.
The red solid (dashed) line shows $\Delta N_{\rm eff}$ for the hidden U(1)$^\prime$ with
(without) massless fermions, where the number of fermions is set to be $N_f = 5$.
As shown in the text, $N_f$ is bounded below $N_f \geq 5$ to satisfy the anomaly cancellation
conditions.
The dot-dashed lines are for larger hidden gauge groups without massless fermions.
The values of $\Lambda_\phi$ leading to ${\rm Br}(h\to V^\prime V^\prime)=20,\,10,\,1\%$
for $2m_\phi>m_h$ are shown by the filled circles on each line,
from left to right, respectively.
}
\label{fig:Neff}
\end{minipage}
\end{center}
\end{figure}

The hidden sector contributes to the effective number of neutrino species according to
the relation (\ref{dnf}).
The overall size of $\Delta N_{\rm eff}$ is determined by $(g_{\ast \nu}/g_{\ast{\rm dec}})^{4/3}$.
In the right panel of Fig. \ref{fig:g}, we illustrate how it changes with $\Lambda_\phi$ when
the hidden sector communicates with the SM sector through the $|\phi|^2|H|^2$ interaction.
One can see that $(g_{\ast \nu}/g_{\ast{\rm dec}})^{4/3}$ ranges from about $0.047$ to $0.065$
for $\Lambda_\phi>1.7$ TeV, where the invisible Higgs decay rate into hidden
gauge bosons is less than 1MeV.
Note that the quantity $(g_{\ast \nu}/g_{\ast{\rm dec}})^{4/3}$ has a mild dependence
on $N_g$ and $N_f$ since the decoupling temperature of dark radiation relies on them.
In Fig. \ref{fig:Neff}, we show how $\Delta N_{\rm eff}$ changes as a function
of $\Lambda_\phi$ for various cases. It is clear that $\Delta N_{\rm eff}\simeq 0.30$ requires
the hidden sector to have $N_g>1$ or massless fermions charged under the hidden gauge group.
The Higgs decay rate into hidden gauge bosons, which is proportional to $N_g/\Lambda^4_\phi$,
is shown by the vertical lines in Fig. \ref{fig:g}, and by filled circles in Fig. \ref{fig:Neff}.

Let us take the minimal hidden gauge group, $G^\prime=$U(1)$^\prime$.
Then the contribution of the massless hidden photon to $N_{\rm eff}$ is smaller than 0.1,
and so, we add $N_f$ chiral fermions $\psi_i$ having a hidden Abelian charge $q^\prime_i$.
Here we define those chiral fermions to be left-handed ones without loss of generality.
The U(1)$^\prime$ charge assignment should be such that vector-like mass terms are forbidden
for $\psi_i$, since otherwise they naturally acquire a heavy mass and do not play an important role
in the present context.
However, the number of hidden fermions and their charge assignment cannot be chosen arbitrarily
because they are subject to two anomaly-free conditions.
One is $\sum_i q^{\prime 3}_i=0$ for the absence of $[{\rm U(1)}^\prime]^3$ anomaly, and the other is
$\sum_i q^\prime_i=0$ to cancel the gravitational $[{\rm U(1)}^\prime]\times[{\rm graviton}]^2$
anomaly.
This problem was systematically studied in Ref.~\cite{Batra:2005rh}, and its connection
to dark matter was studied in Ref.~\cite{Nakayama:2011dj}.
 Under the natural assumption that the hidden Abelian charges are rational numbers,
one can show that the only solutions to the anomaly-free conditions are vector-like fermions
for $N_f\leq 4$ \cite{Batra:2005rh,Nakayama:2011dj}.
Therefore $N_f\geq 5$ is required for the quantum consistency of the gauge
theory.  We can take the charges to be integers as the overall normalization of the U(1)$^\prime$ charges
is arbitrary.  One such example is five chiral fermions carrying U(1)$^\prime$ charges,
$(1,5,-7,-8,9)$.

Using anomaly-free conditions, we find that the hidden sector with unbroken
Abelian gauge group contributes to the effective number of neutrino species as
\bea
\label{nf5}
\Delta N_{\rm eff} \;\simeq\; 0.047 \left(\frac{8}{7}+N_f\right) \geq 0.288,
\eea
for $\Lambda_\phi$ larger than about $10^5$ GeV.
The equality in the last inequality becomes an equality for $N_f = 5$, the smallest number
of massless chiral fermions.
For $\Lambda_\phi$ around $10^3$ GeV, $\Delta N_{\rm eff}$ is increased by a factor
about 1.4.
It is worth noting that $\Delta N_{\rm eff} \simeq 0.29$ is realized in a large portion
of the parameter space.
The only requirement is that the hidden sector fields decouple at a temperature above the
electroweak phase transition.
In particular, the mass of $\phi$ does not need to be finely tuned.
For larger $N_f$, $\Delta N_{\rm eff}$ increases, and so, there is an upper bound on the
number of chiral fermions.
If we double the fermion number as in the case with the $Z_2$ symmetry (see the discussion below),
the contribution to the effective neutrino number is $\Delta N_{\rm eff} \simeq 0.52$
for $N_f = 10$.
If the mass of $\phi$ is sufficiently light, the hidden particles will be decoupled
at lower temperature, contributing a larger amount to $\Delta N_{\rm eff}$.

For the hidden sector with U(1)$^\prime$, we have assumed that the kinetic mixing is vanishingly
small in order to satisfy the experimental bounds~\cite{Davidson:2000hf}.
This can be realized in an extra dimensional set-up where the hidden
gauge field as well as the chiral fermions reside on a hidden brane separated from the SM brane,
while the scalar $\phi$ is in the bulk~\cite{Chen:2008yi}.
It is also possible to assign a $Z_2$ symmetry under which $A'_\mu$ flips the sign so that
the kinetic mixing is forbidden by the symmetry.
The hidden gauge interactions are possible if the chiral fermions as well as the scalar form a doublet,
i.e., $2N_f$ chiral fermions and $2$ scalars. There are also other ways to suppress the kinetic mixings.
See e.g. Refs.~\cite{Dienes:1996zr,Abel:2006qt,Goodsell:2009xc}.

The dark radiation under consideration has an important difference from other conventional
candidates.
In the previous example, the dark radiation is composed of $A'_\mu$ and $\psi_i$, which have unbroken
gauge interactions among them.
That is to say, they are self-interacting dark radiation.\footnote{
As we assume that the hidden sector particles are thermalized through gauge interactions, the equilibrium
in the hidden sector is maintained even after the decoupling from the SM sector.
} As a result,
they have no anisotropic stress, and so the viscosity parameter is zero, while
the sound velocity is given by the standard value as they are massless.\footnote{Here we assume that the gauge interactions
are weak. If they are relatively strong,  there can be sizable corrections to the sound velocity.}
Such self-interacting dark radiation should have left distinct signatures in the CMB angular power spectrum.
On the other hand, if the hidden sector contains only hidden photons for one or several
Abelian factors without massless hidden fermions, one is led to non-interacting dark radiation.

So far we have assumed unbroken U(1)$^\prime$ symmetry.
If $\phi$ acquires a non-zero expectation value, the hidden gauge boson becomes massive.
It depends on the charge assignment whether or not the hidden chiral
fermions remain light. In particular, if the charges of fermions are fractional rational numbers
while the scalar has an unit charge, they can remain massless.\footnote{
Some of them may be massive and contribute to dark matter as in Ref.~\cite{Nakayama:2011dj}.}
The contribution to $\Delta N_{\rm eff}$ can be similarly calculated in this case.
In addition, there appears a mixing between $\phi$ and the SM Higgs boson, and thus
experimental bounds on the scalar coupling should be taken into consideration, but the constraint will
be relaxed for heavier $m_\phi$ and smaller $\lambda$.

Let us continue to discuss another example of self-interacting dark radiation.
Up to this point we have mainly considered an Abelian gauge symmetry, but it is also possible
to consider non-Abelian gauge symmetries~\cite{Nakayama:2010vs}.
A kinetic mixing is then forbidden by the symmetry, and the observational bound is much weaker.
The gauge interaction can be weakly coupled until now if the gauge coupling
is sufficiently small at high energy scales, say, $\alpha^\prime=g^{\prime 2}/4\pi < 10^{-2}$.
The thermalization of the hidden gauge bosons is possible by adding either a scalar as in the previous
example or matter fields which are charged under both
the hidden and the SM gauge symmetries. One of the advantages of non-Abelian symmetry is that
the latter possibility will be viable in a supersymmetric theory as the scalar coupling with the SM
Higgs fields is suppressed by supersymmetry breaking effects. As an example,
let us consider $G^\prime={\rm SU(}N{\rm )}^\prime$.
Then the contribution of massless gauge bosons to the effective neutrino number is
\beq
\Delta N_{\rm eff} \;\simeq\; 0.054\, (N^2-1),
\eeq
which is equal to $0.16$, $0.43$ and $0.80$ for $N=2$, $3$ and $4$.
It is straightforward to add chiral fermions in this case. We can forbid mass terms for those
fermions by imposing a global U(1) symmetry.
It is also possible to combine multiple gauge symmetries and introduce matter fields with
various charge assignments.
In general, the dark radiation could be a mixture of interacting and non-interacting light particles.
Also it is likely that there are various gauge symmetries in the hidden sector, some of which
become strong at low energy scales, while the others remain weakly coupled until now.
The former and matter fields charged under the symmetries may contribute to dark matter,
while the latter contributes to dark radiation.
It is also possible that some of the weakly interacting dark radiation at the recombination
epoch become strongly coupled and massive at later times, contributing to some part of dark matter.
It will be interesting to consider a concrete example where both dark radiation and dark matter
are self-interacting~\cite{Higaki:2013vuv}.

The hidden gauge symmetries may be ubiquitous in the string landscape, and some of them
may be thermalized after inflation, together with the SM sector. It is conceivable that
the inflaton first reheats the hidden gauge sector, and then the SM sector gets thermalized
(or ``recouped") at later times through renormalizable couplings like (\ref{quartic}).
Interestingly, the presence of dark radiation is an inevitable outcome in this scenario.
Such scenario can naturally be embedded into inflation models where the inflaton potential arises
from non-perturbative interactions in the hidden sector and the interactions become weakly coupled
at the inflaton potential minimum.
This can be realized as follows.
Suppose that some of the hidden quarks are heavy during inflation and become massless after inflation.
Then the sign of beta-function for the hidden gauge symmetry can change after inflation,
and the interaction becomes weak at low energy scales.
This is the case if the hidden quarks acquire a mass from the coupling to the inflaton which travels
a large amount during the last $50-60$ e-foldings as in the chaotic inflation
model~\cite{Harigaya:2012pg}, or if the hidden quarks are massive due to a large expectation value
of flat directions~\cite{Jeong:2013xta}.

Lastly we briefly discuss the embedding of the present scenario into a supersymmetric framework.
If we introduce a coupling $|\phi|^2 (|H_u|^2 + |H_d|^2)$ in the K\"ahler potential,
the scalar coupling (\ref{quartic}) is highly suppressed by supersymmetry breaking effects.
Here $H_u$ and $H_d$ are up- and down-type Higgs doublet superfield.
A sizable coupling connecting the hidden sector with the visible sector can be obtained
in a natural way within the next to minimal supersymmetric SM (NMSSM) through
the superpotential terms,
\bea
\Delta W = y S H_uH_d + \kappa S \Phi\bar \Phi,
\eea
where $S$ is the singlet Higgs superfield in the NMSSM sector, and hidden matter fields $\Phi+\bar \Phi$
are vector-like under the hidden gauge group.
The NMSSM maintains most of the nice features of the MSSM while allowing richer physics in
the Higgs and neutralino sectors \cite{NMSSM}.
In addition, the Higgs boson mass around 125.5 GeV can naturally be accommodated owing to the
additional NMSSM contribution from the $SH_uH_d$ term.
The superpotential $\Delta W$ gives rise to the scalar interaction,
\bea
-{\cal L} = y\kappa^* (\phi\bar \phi)^*H_uH_d + {\rm h.c.},
\eea
where $\phi+\bar\phi$ denote the scalar component of $\Phi+\bar\Phi$, and are assumed
to be fixed at the origin so that the hidden gauge symmetry remains unbroken.
The above coupling can be regarded as the supersymmetric analogue of (\ref{quartic}),
and it follows that the hidden sector can be thermalized from the NMSSM sector, or
vice versa.
This set-up has implications not only for the Higgs boson decay into dark radiation
but also for dark matter and inflation model building.
We leave these issues for future work.

In this letter we have considered a simple class of models where the dark radiation has self-interactions and
therefore does not free stream. In the case of U(1)$^\prime$ symmetry with chiral fermions, the contribution to
the effective number of neutrinos is bounded below as (\ref{nf5}). Interestingly, it is equal to about $0.29$
for the smallest number of chiral fermions, $N_f = 5$, free from anomalies.
The self-interacting dark radiation can be probed by the Planck satellite and future CMB observations,
and it may shed light on the hidden sector and its properties.

\section*{Acknowledgment}
FT thanks Tsutomu Yanagida for communication.
This work was supported by
Grant-in-Aid for Scientific Research (C) (No. 23540283) [KSJ],  Scientific Research on Innovative
Areas (No.24111702 [FT], No. 21111006 [FT] , and No.23104008 [KSJ and FT]), Scientific Research (A)
(No. 22244030 and No.21244033) [FT], and JSPS Grant-in-Aid for Young Scientists (B)
(No. 24740135) [FT], and Inoue Foundation for Science [FT].
This work was also supported by World Premier International Center Initiative
(WPI Program), MEXT, Japan [FT].


\begin{thebibliography}{99}
\bibitem{Ade:2013zuv}
  P.~A.~R.~Ade {\it et al.}  [Planck Collaboration],
  arXiv:1303.5076 [astro-ph.CO].

\bibitem{Nakayama:2010vs}
K.~Nakayama, F.~Takahashi and T.~T.~Yanagida,
Phys.\ Lett.\ B {\bf 697} (2011) 275
[arXiv:1010.5693 [hep-ph]].


\bibitem{Ichikawa:2007jv}
K.~Ichikawa, M.~Kawasaki, K.~Nakayama, M.~Senami and F.~Takahashi,
JCAP {\bf 0705} (2007) 008
[arXiv:hep-ph/0703034].

\bibitem{Higaki:2013lra}
  T.~Higaki, K.~Nakayama and F.~Takahashi,
  arXiv:1304.7987 [hep-ph].

\bibitem{Weinberg:2013kea}
  S.~Weinberg,
  arXiv:1305.1971 [astro-ph.CO].

\bibitem{Baek:2013qwa}
  S.~Baek, P.~Ko and W.~-I.~Park,
  JHEP {\bf 1307}, 013 (2013)  [arXiv:1303.4280 [hep-ph]].

\bibitem{Chacko:2003dt}
  Z.~Chacko, L.~J.~Hall, T.~Okui and S.~J.~Oliver,
  Phys.\ Rev.\ D {\bf 70}, 085008 (2004)
  [hep-ph/0312267].

\bibitem{Archidiacono:2013lva}
  M.~Archidiacono, E.~Giusarma, A.~Melchiorri and O.~Mena,
  arXiv:1303.0143 [astro-ph.CO].

\bibitem{Franca:2013zxa}
  U.~Franca, R.~A.~Lineros, J.~Palacio and S.~Pastor,
  arXiv:1303.1776 [astro-ph.CO].

\bibitem{Bell:2005dr}
  N.~F.~Bell, E.~Pierpaoli and K.~Sigurdson,
  Phys.\ Rev.\ D {\bf 73}, 063523 (2006)
  [astro-ph/0511410].


\bibitem{Cirelli:2006kt}
  M.~Cirelli and A.~Strumia,
  JCAP {\bf 0612}, 013 (2006)
  [astro-ph/0607086].


\bibitem{Friedland:2007vv}
  A.~Friedland, K.~M.~Zurek and S.~Bashinsky,
  arXiv:0704.3271 [astro-ph].


\bibitem{Smith:2011es}
  T.~L.~Smith, S.~Das and O.~Zahn,
  Phys.\ Rev.\ D {\bf 85}, 023001 (2012)
  [arXiv:1105.3246 [astro-ph.CO]].


\bibitem{Archidiacono:2011gq}
  M.~Archidiacono, E.~Calabrese and A.~Melchiorri,
  Phys.\ Rev.\ D {\bf 84}, 123008 (2011)
  [arXiv:1109.2767 [astro-ph.CO]].

\bibitem{Diamanti:2012tg}
  R.~Diamanti, E.~Giusarma, O.~Mena, M.~Archidiacono and A.~Melchiorri,
  arXiv:1212.6007 [astro-ph.CO].


\bibitem{Gerbino:2013ova}
  M.~Gerbino, E.~Di Valentino and N.~Said,
  arXiv:1304.7400 [astro-ph.CO].



\bibitem{Davidson:2000hf}
  S.~Davidson, S.~Hannestad and G.~Raffelt,
  JHEP {\bf 0005}, 003 (2000)
  [hep-ph/0001179].


\bibitem{Simon:2012ik}
  F.~Simon,
  arXiv:1211.7242 [hep-ex].  

\bibitem{Giardino:2013bma}
  P.~P.~Giardino, K.~Kannike, I.~Masina, M.~Raidal and A.~Strumia,
  arXiv:1303.3570 [hep-ph].


\bibitem{Batra:2005rh}
  P.~Batra, B.~A.~Dobrescu and D.~Spivak,
  J.\ Math.\ Phys.\  {\bf 47}, 082301 (2006)
  [hep-ph/0510181].

\bibitem{Nakayama:2011dj}
  K.~Nakayama, F.~Takahashi and T.~T.~Yanagida,
  Phys.\ Lett.\ B {\bf 699}, 360 (2011)
  [arXiv:1102.4688 [hep-ph]].

\bibitem{Chen:2008yi}
  C.~-R.~Chen, F.~Takahashi and T.~T.~Yanagida,
  Phys.\ Lett.\ B {\bf 671}, 71 (2009)
  [arXiv:0809.0792 [hep-ph]].

\bibitem{Dienes:1996zr}
  K.~R.~Dienes, C.~F.~Kolda and J.~March-Russell,
  Nucl.\ Phys.\ B {\bf 492}, 104 (1997)
  [hep-ph/9610479].

\bibitem{Abel:2006qt}
  S.~A.~Abel, J.~Jaeckel, V.~V.~Khoze and A.~Ringwald,
  Phys.\ Lett.\ B {\bf 666}, 66 (2008)
  [hep-ph/0608248].

\bibitem{Goodsell:2009xc}
  M.~Goodsell, J.~Jaeckel, J.~Redondo and A.~Ringwald,
  JHEP {\bf 0911}, 027 (2009)
  [arXiv:0909.0515 [hep-ph]].


\bibitem{Higaki:2013vuv}
See e.g.
%
  F.~-Y.~Cyr-Racine and K.~Sigurdson,
  Phys.\ Rev.\ D {\bf 87}, 103515 (2013) [arXiv:1209.5752 [astro-ph.CO]];
%
  T.~Higaki, K.~S.~Jeong and F.~Takahashi,
  arXiv:1302.2516 [hep-ph];
  J.~Fan, A.~Katz, L.~Randall and M.~Reece,
  arXiv:1303.1521 [astro-ph.CO].

\bibitem{Harigaya:2012pg}
  K.~Harigaya, M.~Ibe, K.~Schmitz and T.~T.~Yanagida,
  Phys.\ Lett.\ B {\bf 720}, 125 (2013)
  [arXiv:1211.6241 [hep-ph]].

\bibitem{Jeong:2013xta}
  K.~S.~Jeong and F.~Takahashi,
  arXiv:1304.8131 [hep-ph].

\bibitem{NMSSM}
For a review, see
  M.~Maniatis,
  Int.\ J.\ Mod.\ Phys.\  {\bf A25}, 3505-3602 (2010).
  [arXiv:0906.0777 [hep-ph]];
  U.~Ellwanger, C.~Hugonie, A.~M.~Teixeira,
  Phys.\ Rept.\  {\bf 496}, 1-77 (2010).
  [arXiv:0910.1785 [hep-ph]].



\end{thebibliography}
\end{document}